\documentclass[acmsmall]{acmart}

\AtBeginDocument{%
  \providecommand\BibTeX{{%
    \normalfont B\kern-0.5em{\scshape i\kern-0.25em b}\kern-0.8em\TeX}}}

\acmYear{2020}

\begin{document}

\title[Encounters with Visual Misinformation]{Encounters with Visual Misinformation and Labels Across Platforms}
\subtitle{An Interview and Diary Study to Inform Ecosystem Approaches to Misinformation Interventions}


\author{Emily Saltz}
\email{emily@partnershiponai.org}
\orcid{1234-5678-9012}
\author{Claire Leibowicz}
\authornotemark[1]
\email{claire@partnershiponai.org}
\affiliation{%
  \institution{The Partnership on AI}
  \streetaddress{115 Sansome St, Ste 1200}
  \city{San Francisco}
  \state{CA}
  \postcode{94104}
}

\author{Claire Wardle}
\email{claire@firstdraftnews.com}
\affiliation{%
  \institution{First Draft}
  \streetaddress{219 West 40th Street, 14th floor}
  \city{New York}
  \state{NY}
  \postcode{10018}
}



\begin{abstract}
Since 2016, the amount of academic research with the keyword "misinformation" has more than doubled \cite{dimensions}. This research often focuses on article headlines shown in artificial testing environments, yet misinformation largely spreads through images and video posts shared in highly-personalized platform contexts. A foundation of qualitative research is necessary to begin filling this gap to ensure platforms' visual misinformation interventions are aligned with users’ needs and understanding of information in their personal contexts, across platforms. In two studies, we combined in-depth interviews (n=15) with diary and co-design methods (n=23) to investigate how a broad mix of Americans exposed to misinformation during COVID-19 understand their visual information environments — including encounters with interventions such as Facebook fact-checking labels. Analysis reveals a deep division in user attitudes about platform labeling interventions for visual information which are perceived by many as overly paternalistic, biased, and punitive. Alongside these findings, we discuss our methods as a model for continued independent qualitative research on cross-platform user experiences of misinformation that inform interventions.
\end{abstract}

\ccsdesc[500]{General and reference}
\ccsdesc[500]{Human-centered computing}
\ccsdesc[500]{Human-centered computing~User centered design}
\ccsdesc[300]{Human-centered computing~Participatory design}
\ccsdesc[500]{Human-centered computing~User studies}
\ccsdesc[500]{Information systems~Social networks}
\keywords{visual misinformation, manipulated media, misinformation, news media, social media, platform design}

\begin{teaserfigure}
    \includegraphics[width=\textwidth] {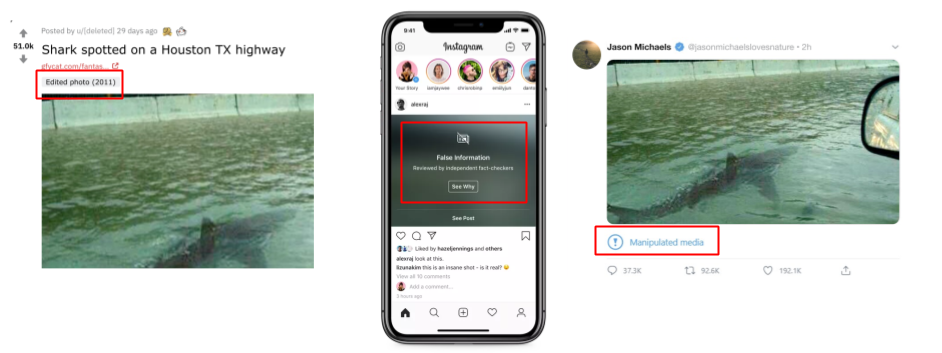}
  \caption{Label stimuli applied to a doctored photo of a shark on a highway \cite{gillin_2017} from left to right: user-applied labels (Reddit), fact-checker overlay labels (Instagram), “Manipulated Media” label without overlay (Twitter).}
  \Description{Three labels with red box highlighting different visual treatments}
\end{teaserfigure}

\maketitle
\section{Introduction}

Whether conspiratorial videos on YouTube, political memes on Instagram, or viral protest photos on Twitter, images and videos on technology platforms have immense power to influence public opinion \cite{mullinix2020feedback, galen-stocking_2020}. As technology platforms find themselves in the position to moderate the credibility of visual content, they are not only tasked with evaluating whether or not content is manipulated and/or misleading, but also with determining how to take action in response to their evaluations. In this case study, we describe why understanding individuals’ encounters and behaviors in response to visual information across platforms is vital for informing effective misinformation interventions.

This need is especially urgent in light of several recent high-profile efforts to label posts, such as Twitter’s labeling of state \cite{support_2020} and manipulated media \cite{synthetic-and-manipulated-media-policy}, and Facebook’s evolving set of labels based on fact-checker ratings \cite{goldshlager_berman_2020}. While such interventions have an intuitive appeal, seeming to be net-positive efforts to fortify media integrity, they take for granted that audiences will accept information from these labels as intended. It is therefore crucial for researchers to independently evaluate the effects of labeling, and other misinformation interventions, in practice.

\subsection{Visual Misinformation}

The unique role of visuals in misinformation is both outsized and understudied. Many people today learn about news on platforms centered on images and videos, such as Instagram, Snapchat, YouTube, and TikTok. In this paper, we define “visual misinformation” as any image or video post with the potential to mislead audiences through false or misleading textual and/or visual elements. Images are also more likely than text to stick in our memory, a phenomenon known as the “picture superiority effect” \cite{defeyter2009picture}. They may also be more likely to influence real world actions: Researchers have found that emotionally-charged positive images like smiling faces, were more likely to affect behaviors contingent upon affective state (e.g., drinking a beverage), compared to equivalently emotional words \cite{winkielman2018influence}. Despite all of this, current misinformation research commonly focuses on text-based properties of information, such as the accuracy of claims in headline and article text \cite{torabi2019big, fazio2020pausing, ecker2014effects, bago2020fake}. While several recent visual misinformation studies do exist, they approach the problem through typological analysis of examples \cite{doi:10.1177/1940161220964780, garimella2020images}, or assessing perceptions of image editing artifacts shown outside of platform contexts \cite{doi:10.1177/1461444818799526}. Though valuable, existing studies lack insight into how people react to these types of visual misinformation examples in their everyday lives.

Outside of academia, evolving tactics for visually misinforming have captured the attention of technology platforms attempting to reduce misinformation’s harm to society \cite{synthetic-and-manipulated-media-policy, bickert_2020, deepfakes_2019}. While much platform and public attention has focused on AI-generated deepfakes \cite{goldshlager_berman_2020}, visual misinformation also includes lower tech “cheapfakes” \cite{paris2019deepfakes} such as miscaptioned images and videos, or edits made using commercial photo and video-editing tools. In the past year, projects like the MediaReview schema \cite{benton_2020} have begun trying to meaningfully describe this range of manipulations — both to the media asset itself and the surrounding components — for use in fact-checker ratings that could be applied to user-facing interventions. 

Unlike platforms like Facebook —  which has written about applying interventions at scale based on “feature-based instance matching” \cite{saltz_noel_leibowicz_wardle_gregory_2020} — we understand visual misinformation as a property of a post’s presentation in a given context, rather than an inherent property of an image or a video asset itself. Even if images or videos themselves are identical, different user contexts can drastically change the artifact’s meaning: consider that even videos or images containing false claims may be shared with accurate user corrections, and “missing context” examples such as unaltered crowd photos \cite{funke_2019} only become visual misinformation if posted alongside inaccurate or misleading claims about what the visuals depict. 

\subsection{Labeling Interventions}

How then are platforms currently moderating varieties of visual misinformation? Alongside content removal or downranking, platforms have increasingly turned to labeling as a strategy to address misinformation. We define "label" here as any textual or visual cue added by to a post with the goal of reducing user belief in misinformation. Labels today take many forms, but consistent detection and accurate application of labeled content remains a challenge. For example, Twitter appended a descriptive label under a tweet that contained incorrect information about California’s absentee ballot process \cite{dwoskin_2020}. At YouTube, despite enacting a policy to label state-sponsored videos, the video-rich platform failed to do so comprehensively \cite{samek_2018, kofman_2019}. 

Yet, even if platforms were able to identify visual misinformation perfectly and apply labels methodically, the question remains: How do current labels affect users' response to misinformation? 

\subsection{Limitations of Current Intervention Research}

Much of the existing academic research on the effects of labeling interventions is based on findings from highly artificial testing conditions, like Mechanical Turk \cite{mason2012conducting}, which lack the ecological validity of platforms themselves \cite{pennycook2020implied, bago2020fake, nyhan2010corrections}. While behavioral research aims to abstract away from complexities of platform environments to unearth underlying psychological principles, we believe that any platform-specific affordances — such as social metrics, comments, personalized feeds and recommendations on social networks — are in fact key elements that influence responses to visual misinformation and its corrections. Changing any of these elements, such as social metrics, has been found to have effects on behaviors related to misinformation, such as the finding from Avram et al. (2020) that, “people are more vulnerable to low-credibility content that shows high levels of social engagement” \cite{avram2020exposure}. While some researchers are trying to investigate misinformation interventions with alternative simulated environments, such as the simulated Facebook intervention from the non-profit activist network Avaaz \cite{avaaz_2020}, these still lack realistic user data and context.

In contrast, while platforms conduct ecologically valid internal research to inform their intervention decisions, they are hesitant to share results of this research beyond high-level blogs in coordination with public relations teams or in closed-door conversations, leaving them to evaluate their own work largely independently. Platforms rarely provide detail on how they conducted their research, or provide metrics for how they analyzed intervention efficacy when reporting publicly. For example, in a recent blog post about warning labels applied to COVID-19 content on Facebook in March 2020, Facebook explained that “when people saw warning labels, 95\% of the time they did not go on to view the original content” with no additional color on methods, effects on belief or engagement, additional metrics and error rates, and other important details signaling the efficacy of such interventions \cite{rosen_2020}. This reality then stifles the evolution of scholarship, testing, and evaluation in the misinformation space that can contribute to more effective interventions. 

Even if a single platform were to conduct ecologically valid research on their interface and chose to share more about their research, the reality is that most users interact with content across platforms \cite{shearer2019americans}. Assessing any single platform in isolation will offer only a partial picture of behavior, given that users consume and share visuals across platforms – as seen in the repeated attempts of users to share the misleading “Plandemic” video about COVID-19 across YouTube, Facebook, Twitter, forums and message groups, despite attempts at coordinated platform removal \cite{dwoskin_2020-2}. 

\subsection{Motivation for this Study}

This case study aims to shed light on how visual information issues play out in practice, to emphasize the limitations of current interventions across platforms, and to identify opportunities for practical recommendations and further research. We present a qualitative methodology for deepening understanding of how people react to visual information across platforms. 

\section{Methods}

We conducted two complementary studies: a 60-minute semi-structured interview study with 15 participants using photo elicitation methods in order to gain insight into attitudes about visual misinformation and existing labels (S1), followed by an 11-day diary study with 23 additional participants to capture the range and prevalence of misinformation labels and media examples that people trust and distrust in the moment across platforms (S2). S2 included a final concept sketching activity that gave participants the opportunity to redesign their platform experiences when encountering media about current events. The visual information discussed in both studies was primarily focused on COVID-19 in order to provide a baseline topic to reference that was likely relevant to all participants. 

Both studies were recruited and conducted in English remotely through dscout, an online qualitative research platform. All participants signed terms of service and consented to each study, with additional consent language written in consultation with dscout project managers and Partnership on AI legal counsel. Semi-structured interview protocols and diary instructions and prompts are provided in the supplementary materials \cite{DVN/QAMZPO_2020}.

\subsubsection{Study Participants}

We aimed to recruit a broad mix of Americans who we believed regularly saw low credibility information on digital platforms, in order to ensure we were talking to people more likely to encounter the types of visual misinformation we were studying. Given the early discovery phase of this research, we prioritized depth of analysis per interview over quantity of interviews in order to better define our target populations for future study. 

In S1, we recruited 15 American English-speaking participants. To screen for participants we believed were regularly exposed to digital misinformation, we relied on self-reported news consumption of at least one low credibility site on the Iffy Index of of Unreliable Sources \cite{iffy.news_2020}, an index which includes only sites with low factual-reporting levels as determined by the Media Bias/Fact Check methodology \cite{mbfc_2020}. 15/15 participants self-reported media consumption from at least one Iffy Index source. Additionally, all participants self-reported seeing and sharing information about COVID-19 on social media at least weekly. Participants included a mix of genders (7 female, 8 male) race/ethnicities (3 Asian, 1 Black, 2 Hispanic, 1 Middle Eastern or North African, 8 White), ages (20–63), locations (6 urban, 9 suburban), income levels (Less than \$25,000-Over \$150,000), education levels (high school graduate to post-graduate college), and politics, skewing Republican (8 Republicans, 4 Democrats, 3 Independents).

S2 featured 23 participants chosen with similar recruiting criteria. Based on findings from S1 that many people had not personally encountered labels, we also added a screening question about past exposure to labels to ensure that all participants had reported seeing labels at least “Sometimes (a handful of times but not regularly)” or more. Additionally, based on observing strong partisan trends in attitudes in S1, we more explicitly sought partisan balance to better understand these differentiators (8 Republicans, 7 Independent/Other, 8 Democrats). Participants were a mix of genders (11 female, 12 male), race/ethnicities (4 Asian, 6 Black, 4 Hispanic, 1 Middle Eastern or North African, 8 White), ages (20-63), locations (9 urban, 13 suburban, 1 rural), income (Less than \$25,000-Over \$150,000) and education levels (high school graduate to post-graduate college). 23/23 self-reported media consumption from one or more Iffy Index source. See supplementary materials \cite{DVN/QAMZPO_2020} for more participant details. 

\subsubsection{Interview Study Procedure (S1)}


S1 aimed to understand how people understand existing terminology for visual misinformation and labeling interventions. The 60-minute interviews were conducted in four parts. We began with a semi-structured interview about users’ media habits and consumption of COVID-19 information online and offline, and their perceived access to credible vs. non-credible information (A1). Next, interviews featured a think-aloud discussion and comparison of images of existing manipulated media interventions from Instagram, Reddit, and Twitter shown in sequence in a Google slideshow (A2), all as applied to an edited image of the highway shark shown in Figure 1. 

We continued with a series of label terminology and photo/video elicitation tasks for five fact-checked image and video examples corresponding to distinct manipulation types as defined by the MediaReview schema \cite{benton_2020} (A3), as shown in Figure 2. We randomized examples in Google Slides using the Slides Randomizer add-on in order to control for ordering effects. This exercise involved think-aloud discussion about each post and the participant’s ideal intervention and terminology. Examples were shown alongside lightly-edited, original fact-checker claims descriptions, alongside a set of terminology options relevant to each rating type. Finally, interviews ended with open-ended questions to understand how participants feel about platforms labeling media, and why (A4).

To analyze interview data, a moderator and a notetaker debriefed bullet-point notes by interview section and summarized key takeaways after each interview. We then worked through these notes and transcriptions of approximately 15 hours of audio to synthesize findings through bottom-up thematic analysis \cite{braun2006using} to iteratively generate codes for various interview notes and utterances, grouping these codes into successively higher-level themes concerning attitudes toward visual information and labeling interventions. Key themes are presented alongside findings from S2 in Findings and Discussion below.

\subsubsection{Diary Study Procedure (S2)}

\begin{table}
  \caption{Diary Activities: Screenshots of moments (B2) and concept sketching artifacts (B3) can be found in supplementary materials \cite{DVN/QAMZPO_2020}. Examples can also be found later in this document.}
  \label{tab:freq}
  \begin{tabular}{ccl}
    \toprule
    Day Count & Activity & Entry Count\\
    \midrule
    \ 2 & Background (B1) & 1 \\
    \ 7 & Critical image/video moments (B2*) & $\geq$5 \\
    \ 2 & Reflection and concept sketching (B3*) & 1 \\
    \bottomrule
  \end{tabular}
\end{table}

The diary study aimed to build upon and complement findings from S1 by capturing critical moments driving trust and distrust in images and videos in context of their everyday use across platforms, and what are users’ mental models of how the platforms they use moderate visual information and apply interventions, in the context of their own feeds?

To capture in-context experiences, we used dscout's mobile diary tool. The diary began with one entry of background questions in survey format similar to A1 in S1 in order to better socially situate participants and understand their attitudes about visual misinformation and platform moderation (B1). In the next section, spanning seven days, participants were asked to log at least five screenshots and selfie-style videos describing personal encounters with a range of visual information in real time (B2). We used two triggers for logging images or videos from their personal contexts: 1) images and videos that elicited a strong reaction or 2), any time they saw a label applied by a platform like Facebook, Instagram, or Twitter, such as a "false information" or "manipulated media" labels. 

We used “strong reaction” as a trigger to ensure we captured a range of posts that users strongly supported, questioned, or opposed, in order to gain a better holistic sense of the meaningful visual influences in their lives for learning about current events. We also intentionally did not specify a platform, in order to encourage participants to log media wherever they came across it in their lives and get a better sense of the range and fluidity of important media touchpoints in their life. Overall, we captured 121 media entries total, with an average of five entries, ranging from five to seven entries per participant. Of these media entries, 14 contained an example of labeled media: 13/14 from Facebook or Instagram, and one from YouTube.

Finally, the diary ended with participants reflecting on how the images and videos logged over the week related to their broader visual information experiences. They also selected an issue about visual information and sketched and described an alternative user experience for their platform(s) of choice (B3). Preliminary observations of the media entries and sketches are presented to complement and build on key themes from S1 in Findings and Discussion.

\section{Findings and Discussion}

\begin{figure}
    \centering
    \includegraphics[width=0.9\textwidth]{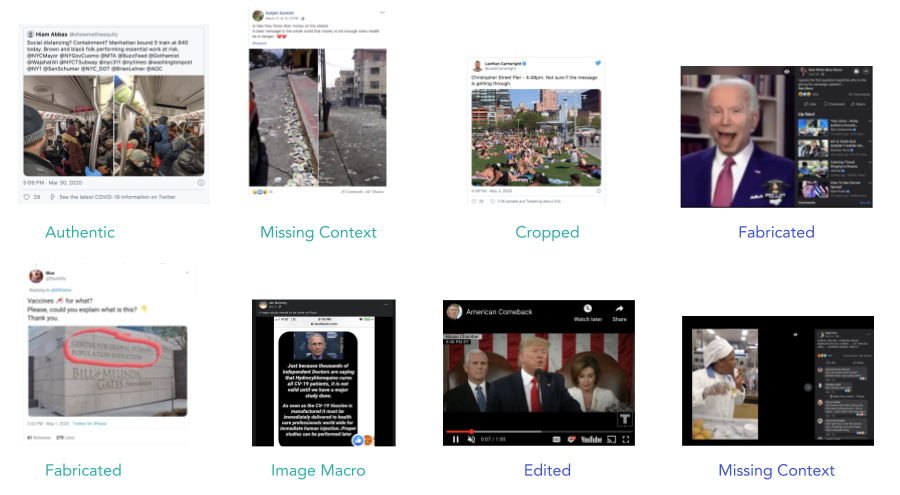}
    \caption{Images (green) and videos (blue) shown to participants in A3 of S1 with categories based on MediaReview\cite{benton_2020} fact-checker ratings.}
    \Description{The images were pulled from the following ratings: Authentic, Missing Context, Fabricated, Cropped, and ImageMacro based on a set of COVID-19-related examples of fact-checker ratings in MediaReview. The three videos were pulled from the Edited, Fabricated, Missing Context MediaReview ratings.}
    \label{fig:types}
\end{figure}

Preliminary findings from both studies show deeply divided attitudes toward platform interventions: Some experience platform interventions as patronizing and political, targeting their preferred media and autonomy to understand content for themselves; others see interventions like labels as positive, if insufficient steps, to clarify media credibility, especially for who they perceived as more gullible others. 

\subsection{Information Ecosystem}
\subsubsection{It’s the Message, not the Media Format: Instances of visual misinformation as part of larger issues in divisive public discourse}
In our initial interviews, many participants described feeling exhausted by an overwhelming deluge of conflicting, negative, emotionally-charged content on social media (S1-P06, S1-P11, S1-P15). Reactions to misleading and false image and video posts were just manifestations of broader feelings of discontent about the polarized state of news media and current events – in other words, media format was a less salient factor than the narrative and sociocultural framing of the post (e.g. pro-mask vs. anti-mask). Several people observed how stories sometimes contradicted each other, amplifying feelings of uncertainty:

\begin{quote}
 \textit{“You hear one news station say one thing [about COVID-19] and one news station saying another. A lot of times it’s political...I don't really know who to trust.”} –S1-P03
\end{quote}

Participants in B1, S2 echoed these sentiments. When asked to provide “three adjectives that best describe how digital platforms’ handling of image and video posts about current news events such as COVID-19 [made them] feel,” the most frequent words included: Anxious(4), Uneasy(3), Confused(3), and Frustrated(3). 

\subsubsection{Group Chats as Safe Havens, Until They’re Not}
Many were disillusioned by contentious comments sections on public platforms like Facebook in particular, and avoided engaging for risk of being misunderstood and getting into online fights that might affect their relationships (S1-P06, S1-P08, S1-P09, S1-P14). In response, some preferred sharing posts to smaller, trusted networks where they felt more safe. Despite this, the family or friend group chat also figured into many participants’ stories of media that led to personal conflict in their lives, as in one woman’s exchange with a friend about masks:

\begin{quote}
 \textit{“I jokingly sent out this short video of masks being made in some third world country [to my lifelong friend who works for a hospital]. She got very upset that I had sent this out to our [WhatsApp] friend group. And I was like, oh wow, I really didn't even put that much thought into it. It was just like, I didn't know [if] this was a real video. And am I going to stop wearing masks? No. But so I was kind of a little taken aback that I had to explain myself. You know, we're 50-plus-years friends. You have to know that I don't think of these people as just numbers.”} –S1-P09
\end{quote}

\subsubsection{Gap Between Perceived and Experienced Visual Misinformation Issues}

In perceptions of visual information in particular, there was a gap between simplified mental models and expectations of issues and issues experienced in real time. In A1, S2, participants answered that they felt tactics like image/video editing caused the most harm (including concerns about cropping from 13/23, selective video editing by 10/23, and "photoshopped" and deepfake edits by 9/23), compared to grounded media moments submitted during diaries, which exposed that the most common “harmful” issues related to disagreements in narrative framing: such as miscontextualized media, or media they understood as accurate but was discounted as false (affecting 9/23 and 8/23 participants respectively).

It’s from this news environment that labels on image and video posts are seen by many as further aggravating divisions: Some expect clarity from platforms through more aggressive labeling and removal of misleading content (eight participants). Others search for clarity through the crowd and/or a mix of mainstream and alternative sources, distrusting platforms to objectively rate media (seven participants). This division was further validated in context by diary entries containing labels.

\subsection{Label Attitude A: Platforms are responsible for fact-checking}

Many believe it’s the platform’s responsibility to be clear and objective about what’s true or false (8/15 in S1), saying: “If it's clearly fake, mark it fake for others” (S1-P05); “I’m happy [labeling] is starting to happen. People see something online and take it as gospel. Platforms need to take social responsibility” (S1-P14); “I feel [labels] would help people understand that not everything that you see and watch is true” (S1-P10). Several diary entries of labels expressed this sentiment, including one Republican participant who encountered a labeled Tucker Carlson video on Facebook:

\begin{quote}
 \textit{"It was really surprising to see something like a huge personality like [Tucker Carlson] would release something that is really not true…I really felt that I had to read more about it. I am glad Facebook marked it as false."} –S2-P22
\end{quote}

\subsubsection{Stronger labels for stronger harms}
Many pro-label participants desired stronger labels for stronger perceived harms, such as those envisioned from a photo of the Bill and Melinda Gates Center (as shown in Figure 2) which edited to read "Center for Global Human Population Reduction:” “Once something is out there, it’s hard to unsee it” (S1-P07). “[This manipulated image is] trying to tear [Bill Gates] down. There's a lot of fake images online that make you want to look, feel an emotion. But this one goes farther in the harm it's trying to do, so a stronger word like ‘Fabricated’ is needed.” “If you just have a label, can be easy to miss, and also most people trying to elicit emotional response and not examining the image more after feeling that way” (S1-P15). “Someone scrolling through their feed might see this picture, and if it's not obscured and they don’t take the time to read the verification, the damage is already done” (S1-P14).

\subsubsection{If it’s ‘fake,’ remove it}

One attitude in the diary entries, expressed by only one person in our initial interviews (S1-P05), was that removal would sometimes be preferable to labeling. Of 10 entries rated by nine participants as “completely inaccurate,” in a majority of instances (6/10), wanted the media removed. For example, in response to a conservative Instagram meme with a false claim that George Floyd died of a drug overdose: 

\begin{quote}
 \textit{“If Instagram knows that these claims are just completely false, it needs to be taken off the app. Why would you allow something that is incorrect to keep spreading? Just because you put a warning doesn’t mean people who already subscribe to that ideology aren’t going to believe it.}” –S2-P01
\end{quote}

\begin{figure}
    \centering
    \includegraphics[width=0.3\textwidth]{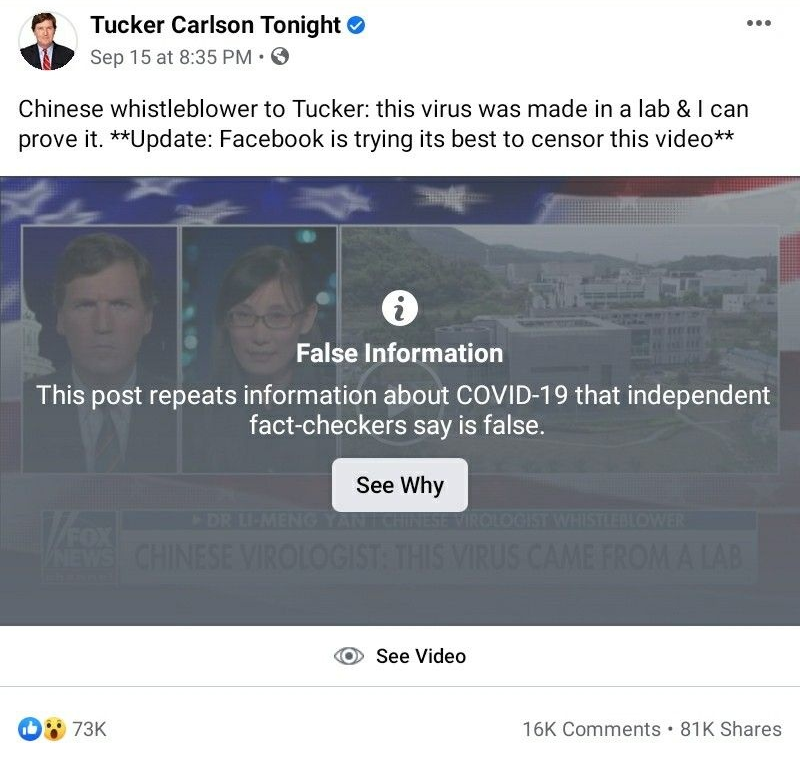}
    \caption{Screenshot of a fact-checking label applied to a video posted on Facebook by Tucker Carlson, as encountered by S2-P22.}
    \Description{}
    \label{fig:taze}
\end{figure}

\subsection{Label Attitude B: Platforms are policing free speech with bias}

In contrast, many viewed labeling (7/15 in S1) as judgmental, paternalistic, and against the platform ethos. There were comparisons to platforms as police: “I don't want Instagram and Twitter to be the police. It's on me to understand the facts within these posts...would they still be doing this if someone else was in the White House?” (S1-P08). This was seen by some as at odds with the intent of social media: “It's an open forum, therefore it's its nature. Posts on Reddit should not be fact-checked” (S1-P12). 

This hostile attitude toward labeling was also seen in the diary study, as one participant encountered a post that he independently investigated and determined was accurate as originally posted:

\begin{quote}
 \textit{“When I see fact-checkers on a post, it agitates me. Because they think [they] know it all. The original post was correct. I’ve seen a lot of fact-checking done [based] on personal opinion, using fake articles to support their agenda."} –S2-P16
\end{quote}

\subsubsection{False positives: Label severity mismatched with harm}

Others also noted times they felt labels were inappropriately applied without taking into account the poster’s intent. One participant, S1-P04, described two times she felt she had personally had labels applied to her posts without merit: The first post was a meme that she felt “was very obviously a joke.” In response to the label, she said she was “livid.” The second time she knowingly shared a false post, but it was in order to educate others: “I had actually in my post shared this picture and in my own words said ‘this is why you need to fact-check.’ And they still blocked it!...It's like they look at the pictures or the headline that you're posting and not look at what you post with it.” 

Though she maintained that neither post was intended to mislead, she ended up deleting both, claiming “[the label] makes you look like an idiot.” Both experiences reduced her trust in the accuracy of fact-checking labels applied by Facebook. Another participant, S2-P09, compared these “false positive” labels to flaws in the criminal justice process: “People I know post things and [get labeled] and it’s almost like the death penalty: Sometimes you kill an innocent person.” 

\subsubsection{Overlays as censorship}

Some felt the overlay style could be particularly punitive and patronizing, described by several participants as “offensive” or “censorship,” saying: “Instagram is ‘the most offensive’...people are smart enough to understand whether [the image in figure 1 is] a shark or not (S1-P12); “Why do I need somebody to tell me that something's false? Why not just let me find out if it's false or not. I don't want somebody to tell me. Like, so my news is being censored? It’s so offensive.” (S1-P09) “I want to decide for myself whether the info being blocked is true or false through my own lens” (S1-P11).

\subsubsection{The mythical “unbiased” label}

Many expressed the desire for a “common ground” (S1-P15) neutral authority or third-party that all would trust, rather than platforms seen as profit-driven and politically-biased, picking and choosing what to check. For some, fact-checkers themselves were seen as biased: “A lot of the stuff I listen to on YouTube is constantly being shut down. And the fact checkers are who? It's really a moral fact checker, it's not really facts. People are either closing down platforms or are silencing voices in some way because they don't like what's being said...What’s their agenda? Who’s paying them?” (S1-P09). 

What is "neutral?" Several wished for automated interventions that could be ostensibly applied without personal bias across all content. Some suggested, “Have the algorithm take care of it” (S1-P06), or “[I like Twitter’s label over Facebook’s because] it feels like it's less like Big Brother is watching...it's like identified by a machine rather than some human checkers" –S1-P05. Others felt neutrality could be accomplished if a story was corroborated by news outlets with diverse perspectives: “If it’s saying the same on the left and right I know I can trust it” –S1-P12.

\subsection{Design Opportunities}

Beyond current labels, what other forms might interventions take? The following aren’t intended as recommendations to be implemented directly, but rather offer insight into how participants understand the problem space. Drawings in B3 of S2 validated that participants largely held a simplified “true” vs. “false” mental model of misinformation, and wished for moderation more aligned with their definitions of credibility applied consistently across platform feeds. 

S2 participants sketched concepts in response to the visual misinformation issues and platforms they felt were most important in their own lives. Most built on strategies currently employed by social media platforms, such as: better fact-checks, including additional source information and ability to check at will, across platforms (9 responses) and better filters, including filtering only for confirmed “verified” news media (6 responses). Notably, both of these strategies locate the problem as a failure of platforms to accurately categorize information as “true” and/or “false.”

Additional strategies included: centering user perspectives through crowdsourcing (3 responses), editing feeds to include less negative, divisive media overall (2 responses), or improving platform appeal processes (2 responses), and external oversight (1 response) processes.

\section{Conclusion and Future Work}

These findings have profound implications for platform interventions. Though drawing from a limited sample size, both studies vividly demonstrate the intense emotional reactions evoked by labels for misinformation in general – perceived by many as overly paternalistic, biased, and punitive. As such, current post-level labeling interventions seem likely to continue to provoke feelings of distrust and hostility toward platforms and content correction labels. At a minimum, these insights should push platforms, researchers, and journalists to more holistically understand the feelings of discontent and uncertainty that lead many to engage with misinformation in alternative media narratives in the first place, and invest in ways to legitimately earn the trust of consumers disillusioned with these institutions.

These findings also complicate discussion around “the backfire effect” — the idea that “when a claim aligns with someone’s ideological beliefs, telling them that it’s wrong will actually make them believe it even more strongly” \cite{swire2020searching}. Though this phenomenon is thought to be rare, our findings suggest that emotionally-charged, defensive backfire reactions may be common in practice for American social media users encountering corrections on social media posts about news topics. While our sample size was too small to definitively measure whether the labels actually strengthened beliefs in inaccurate claims, at the very least, reactions described above showed doubt and distrust toward the credibility of labels – often with reason, as in the case of “false positive” automated application of labels in inappropriate contexts. Any additional insight that platforms can provide about the extent of label skepticism and disbelief toward various label contents from their own data will further the ability of independent researchers to make appropriate intervention recommendations. 

As an immediate next step, we aim to evaluate the prevalence and generality of these themes by conducting a demographically representative survey on a broader population. A survey could serve as a supplement to initial qualitative studies, assessing emerging themes from interviews and diary submissions. Additionally, we aim to learn more about global risks not captured in our American interviews through consultation with small focus groups of civil society partner organizations in order to better connect with affected and marginalized communities that are distinctly impacted by harmful online content.

In summary, our findings point to major limitations in current visual misinformation labeling approaches as experienced by end-users. Yet there are fruitful opportunities for any independent researcher to use qualitative methods to work closely with most affected platform users to define context-specific needs. Doing so will bolster creation of emotionally resonant interventions for visual misinformation and information at large. 

\begin{acks}
Thanks to Victoria Kwan, Tommy Shane, Laura Garcia, and Pedro Noel at First Draft, and Stephen Adler, Nicholas Anway, Kemi Bello, Jeffrey Brown, B Cavello, Riccardo Fogliato, Adam Schetky, Jonathan Stray, Tina Park and all of our colleagues at the Partnership on AI, the AI \& Media Integrity Steering Committee, and The Duke Reporters' Lab.
\end{acks}

\bibliographystyle{ACM-Reference-Format}
\bibliography{base}


\begin{thebibliography}{38}


\ifx \showCODEN    \undefined \def \showCODEN     #1{\unskip}     \fi
\ifx \showDOI      \undefined \def \showDOI       #1{#1}\fi
\ifx \showISBNx    \undefined \def \showISBNx     #1{\unskip}     \fi
\ifx \showISBNxiii \undefined \def \showISBNxiii  #1{\unskip}     \fi
\ifx \showISSN     \undefined \def \showISSN      #1{\unskip}     \fi
\ifx \showLCCN     \undefined \def \showLCCN      #1{\unskip}     \fi
\ifx \shownote     \undefined \def \shownote      #1{#1}          \fi
\ifx \showarticletitle \undefined \def \showarticletitle #1{#1}   \fi
\ifx \showURL      \undefined \def \showURL       {\relax}        \fi
\providecommand\bibfield[2]{#2}
\providecommand\bibinfo[2]{#2}
\providecommand\natexlab[1]{#1}
\providecommand\showeprint[2][]{arXiv:#2}

\bibitem[\protect\citeauthoryear{??}{syn}{[n.d.]}]%
        {synthetic-and-manipulated-media-policy}
 \bibinfo{year}{[n.d.]}\natexlab{}.
\newblock
\newblock
\urldef\tempurl%
\url{https://help.twitter.com/en/rules-and-policies/manipulated-media}
\showURL{%
\tempurl}


\bibitem[\protect\citeauthoryear{??}{dim}{[n.d.]}]%
        {dimensions}
 \bibinfo{year}{[n.d.]}\natexlab{}.
\newblock \bibinfo{title}{The next evolution in linked scholarly information}.
\newblock
\newblock
\urldef\tempurl%
\url{https://app.dimensions.ai/discover/publication?search_mode=content&search_text=misinformation&search_type=kws&search_field=full_search}
\showURL{%
\tempurl}


\bibitem[\protect\citeauthoryear{??}{dee}{2019}]%
        {deepfakes_2019}
 \bibinfo{year}{2019}\natexlab{}.
\newblock
\newblock
\urldef\tempurl%
\url{https://ai.facebook.com/blog/deepfake-detection-challenge/}
\showURL{%
\tempurl}


\bibitem[\protect\citeauthoryear{??}{iff}{2020}]%
        {iffy.news_2020}
 \bibinfo{year}{2020}\natexlab{}.
\newblock
\newblock
\urldef\tempurl%
\url{https://iffy.news/}
\showURL{%
\tempurl}


\bibitem[\protect\citeauthoryear{??}{ava}{2020}]%
        {avaaz_2020}
 \bibinfo{year}{2020}\natexlab{}.
\newblock \bibinfo{title}{How Facebook can Flatten the Curve of the Coronavirus
  Infodemic}.
\newblock
\newblock
\urldef\tempurl%
\url{https://secure.avaaz.org/campaign/en/facebook_coronavirus_misinformation/}
\showURL{%
\tempurl}


\bibitem[\protect\citeauthoryear{??}{mbf}{2020}]%
        {mbfc_2020}
 \bibinfo{year}{2020}\natexlab{}.
\newblock \bibinfo{title}{Methodology}.
\newblock
\newblock
\urldef\tempurl%
\url{https://mediabiasfactcheck.com/methodology/}
\showURL{%
\tempurl}


\bibitem[\protect\citeauthoryear{Avram, Micallef, Patil, and Menczer}{Avram
  et~al\mbox{.}}{2020}]%
        {avram2020exposure}
\bibfield{author}{\bibinfo{person}{Mihai Avram}, \bibinfo{person}{Nicholas
  Micallef}, \bibinfo{person}{Sameer Patil}, {and} \bibinfo{person}{Filippo
  Menczer}.} \bibinfo{year}{2020}\natexlab{}.
\newblock \showarticletitle{Exposure to Social Engagement Metrics Increases
  Vulnerability to Misinformation}.
\newblock \bibinfo{journal}{\emph{arXiv preprint arXiv:2005.04682}}
  (\bibinfo{year}{2020}).
\newblock


\bibitem[\protect\citeauthoryear{Bago, Rand, and Pennycook}{Bago
  et~al\mbox{.}}{2020}]%
        {bago2020fake}
\bibfield{author}{\bibinfo{person}{Bence Bago}, \bibinfo{person}{David~G Rand},
  {and} \bibinfo{person}{Gordon Pennycook}.} \bibinfo{year}{2020}\natexlab{}.
\newblock \showarticletitle{Fake news, fast and slow: Deliberation reduces
  belief in false (but not true) news headlines.}
\newblock \bibinfo{journal}{\emph{Journal of experimental psychology: general}}
  (\bibinfo{year}{2020}).
\newblock


\bibitem[\protect\citeauthoryear{Benton}{Benton}{2020}]%
        {benton_2020}
\bibfield{author}{\bibinfo{person}{Joshua Benton}.}
  \bibinfo{year}{2020}\natexlab{}.
\newblock \bibinfo{title}{Is this video "missing context," "transformed," or
  "edited"? This effort wants to standardize how we categorize visual
  misinformation}.
\newblock
\newblock
\urldef\tempurl%
\url{https://www.niemanlab.org/2020/01/is-this-video-missing-context-transformed-or-edited-this-effort-wants-to-standardize-how-we-categorize-visual-misinformation/}
\showURL{%
\tempurl}


\bibitem[\protect\citeauthoryear{Bickert}{Bickert}{2020}]%
        {bickert_2020}
\bibfield{author}{\bibinfo{person}{Monika Bickert}.}
  \bibinfo{year}{2020}\natexlab{}.
\newblock \bibinfo{title}{Enforcing Against Manipulated Media}.
\newblock
\newblock
\urldef\tempurl%
\url{https://about.fb.com/news/2020/01/enforcing-against-manipulated-media/}
\showURL{%
\tempurl}


\bibitem[\protect\citeauthoryear{Braun and Clarke}{Braun and Clarke}{2006}]%
        {braun2006using}
\bibfield{author}{\bibinfo{person}{Virginia Braun} {and}
  \bibinfo{person}{Victoria Clarke}.} \bibinfo{year}{2006}\natexlab{}.
\newblock \showarticletitle{Using thematic analysis in psychology}.
\newblock \bibinfo{journal}{\emph{Qualitative research in psychology}}
  \bibinfo{volume}{3}, \bibinfo{number}{2} (\bibinfo{year}{2006}),
  \bibinfo{pages}{77--101}.
\newblock


\bibitem[\protect\citeauthoryear{Brennen, Simon, and Nielsen}{Brennen
  et~al\mbox{.}}{0}]%
        {doi:10.1177/1940161220964780}
\bibfield{author}{\bibinfo{person}{J.~Scott Brennen}, \bibinfo{person}{Felix~M.
  Simon}, {and} \bibinfo{person}{Rasmus~Kleis Nielsen}.}
  \bibinfo{year}{0}\natexlab{}.
\newblock \showarticletitle{Beyond (Mis)Representation: Visuals in COVID-19
  Misinformation}.
\newblock \bibinfo{journal}{\emph{The International Journal of Press/Politics}}
  \bibinfo{volume}{0}, \bibinfo{number}{0} (\bibinfo{year}{0}),
  \bibinfo{pages}{1940161220964780}.
\newblock
\urldef\tempurl%
\url{https://doi.org/10.1177/1940161220964780}
\showDOI{\tempurl}
\showeprint{https://doi.org/10.1177/1940161220964780}


\bibitem[\protect\citeauthoryear{Defeyter, Russo, and McPartlin}{Defeyter
  et~al\mbox{.}}{2009}]%
        {defeyter2009picture}
\bibfield{author}{\bibinfo{person}{Margaret~Anne Defeyter},
  \bibinfo{person}{Riccardo Russo}, {and} \bibinfo{person}{Pamela~Louise
  McPartlin}.} \bibinfo{year}{2009}\natexlab{}.
\newblock \showarticletitle{The picture superiority effect in recognition
  memory: A developmental study using the response signal procedure}.
\newblock \bibinfo{journal}{\emph{Cognitive Development}} \bibinfo{volume}{24},
  \bibinfo{number}{3} (\bibinfo{year}{2009}), \bibinfo{pages}{265--273}.
\newblock


\bibitem[\protect\citeauthoryear{Dwoskin}{Dwoskin}{2020a}]%
        {dwoskin_2020}
\bibfield{author}{\bibinfo{person}{Elizabeth Dwoskin}.}
  \bibinfo{year}{2020}\natexlab{a}.
\newblock \bibinfo{title}{Misinformation about coronavirus finds new avenues on
  unexpected sites}.
\newblock
\newblock
\urldef\tempurl%
\url{https://www.washingtonpost.com/technology/2020/05/20/misinformation-coronavirus-plandemic-workaround/}
\showURL{%
\tempurl}


\bibitem[\protect\citeauthoryear{Dwoskin}{Dwoskin}{2020b}]%
        {dwoskin_2020-2}
\bibfield{author}{\bibinfo{person}{Elizabeth Dwoskin}.}
  \bibinfo{year}{2020}\natexlab{b}.
\newblock \bibinfo{title}{Twitter labels Trump's tweets with a fact check for
  the first time}.
\newblock
\newblock
\urldef\tempurl%
\url{https://www.washingtonpost.com/technology/2020/05/26/trump-twitter-label-fact-check/}
\showURL{%
\tempurl}


\bibitem[\protect\citeauthoryear{Ecker, Lewandowsky, Chang, and Pillai}{Ecker
  et~al\mbox{.}}{2014}]%
        {ecker2014effects}
\bibfield{author}{\bibinfo{person}{Ullrich~KH Ecker}, \bibinfo{person}{Stephan
  Lewandowsky}, \bibinfo{person}{Ee~Pin Chang}, {and} \bibinfo{person}{Rekha
  Pillai}.} \bibinfo{year}{2014}\natexlab{}.
\newblock \showarticletitle{The effects of subtle misinformation in news
  headlines.}
\newblock \bibinfo{journal}{\emph{Journal of experimental psychology: applied}}
  \bibinfo{volume}{20}, \bibinfo{number}{4} (\bibinfo{year}{2014}),
  \bibinfo{pages}{323}.
\newblock


\bibitem[\protect\citeauthoryear{Fazio}{Fazio}{2020}]%
        {fazio2020pausing}
\bibfield{author}{\bibinfo{person}{Lisa Fazio}.}
  \bibinfo{year}{2020}\natexlab{}.
\newblock \showarticletitle{Pausing to consider why a headline is true or false
  can help reduce the sharing of false news}.
\newblock \bibinfo{journal}{\emph{Harvard Kennedy School Misinformation
  Review}} \bibinfo{volume}{1}, \bibinfo{number}{2} (\bibinfo{year}{2020}).
\newblock


\bibitem[\protect\citeauthoryear{Funke}{Funke}{2019}]%
        {funke_2019}
\bibfield{author}{\bibinfo{person}{Daniel Funke}.}
  \bibinfo{year}{2019}\natexlab{}.
\newblock \bibinfo{title}{Who needs deepfakes when bogus crowd photos get
  thousands of shares on Facebook?}
\newblock
\newblock
\urldef\tempurl%
\url{https://www.poynter.org/fact-checking/2019/who-needs-deepfakes-when-bogus-crowd-photos-get-thousands-of-shares-on-facebook/}
\showURL{%
\tempurl}


\bibitem[\protect\citeauthoryear{Galen~Stocking}{Galen~Stocking}{2020}]%
        {galen-stocking_2020}
\bibfield{author}{\bibinfo{person}{Patrick van~Kessel Galen~Stocking}.}
  \bibinfo{year}{2020}\natexlab{}.
\newblock \bibinfo{title}{Many Americans Get News on YouTube, Where News
  Organizations and Independent Producers Thrive Side by Side}.
\newblock
\newblock
\urldef\tempurl%
\url{https://www.journalism.org/2020/09/28/many-americans-get-news-on-youtube-where-news-organizations-and-independent-producers-thrive-side-by-side/}
\showURL{%
\tempurl}


\bibitem[\protect\citeauthoryear{Garimella and Eckles}{Garimella and
  Eckles}{2020}]%
        {garimella2020images}
\bibfield{author}{\bibinfo{person}{Kiran Garimella} {and} \bibinfo{person}{Dean
  Eckles}.} \bibinfo{year}{2020}\natexlab{}.
\newblock \showarticletitle{Images and Misinformation in Political Groups:
  Evidence from WhatsApp in India}.
\newblock \bibinfo{journal}{\emph{arXiv preprint arXiv:2005.09784}}
  (\bibinfo{year}{2020}).
\newblock


\bibitem[\protect\citeauthoryear{Gillin}{Gillin}{2017}]%
        {gillin_2017}
\bibfield{author}{\bibinfo{person}{Joshua Gillin}.}
  \bibinfo{year}{2017}\natexlab{}.
\newblock \bibinfo{title}{There are no sharks swimming in the streets of
  Houston or anywhere else}.
\newblock
\newblock
\urldef\tempurl%
\url{https://www.politifact.com/factchecks/2017/aug/28/blog-posting/there-are-no-sharks-swimming-streets-houston-or-an/}
\showURL{%
\tempurl}


\bibitem[\protect\citeauthoryear{Goldshlager and Berman}{Goldshlager and
  Berman}{2020}]%
        {goldshlager_berman_2020}
\bibfield{author}{\bibinfo{person}{Keren Goldshlager} {and}
  \bibinfo{person}{Aaron Berman}.} \bibinfo{year}{2020}\natexlab{}.
\newblock \bibinfo{title}{New Ratings for Fact-Checking Partners}.
\newblock
\newblock
\urldef\tempurl%
\url{https://www.facebook.com/journalismproject/programs/third-party-fact-checking/new-ratings}
\showURL{%
\tempurl}


\bibitem[\protect\citeauthoryear{Kofman}{Kofman}{2019}]%
        {kofman_2019}
\bibfield{author}{\bibinfo{person}{Ava Kofman}.}
  \bibinfo{year}{2019}\natexlab{}.
\newblock \bibinfo{title}{YouTube Promised to Label State-Sponsored Videos But
  Doesn't Always Do So}.
\newblock
\newblock
\urldef\tempurl%
\url{https://www.propublica.org/article/youtube-promised-to-label-state-sponsored-videos-but-doesnt-always-do-so}
\showURL{%
\tempurl}


\bibitem[\protect\citeauthoryear{Mason and Suri}{Mason and Suri}{2012}]%
        {mason2012conducting}
\bibfield{author}{\bibinfo{person}{Winter Mason} {and}
  \bibinfo{person}{Siddharth Suri}.} \bibinfo{year}{2012}\natexlab{}.
\newblock \showarticletitle{Conducting behavioral research on Amazon’s
  Mechanical Turk}.
\newblock \bibinfo{journal}{\emph{Behavior research methods}}
  \bibinfo{volume}{44}, \bibinfo{number}{1} (\bibinfo{year}{2012}),
  \bibinfo{pages}{1--23}.
\newblock


\bibitem[\protect\citeauthoryear{Mullinix, Bolsen, and Norris}{Mullinix
  et~al\mbox{.}}{2020}]%
        {mullinix2020feedback}
\bibfield{author}{\bibinfo{person}{Kevin~J Mullinix}, \bibinfo{person}{Toby
  Bolsen}, {and} \bibinfo{person}{Robert~J Norris}.}
  \bibinfo{year}{2020}\natexlab{}.
\newblock \showarticletitle{The Feedback Effects of Controversial Police Use of
  Force}.
\newblock \bibinfo{journal}{\emph{Political Behavior}} (\bibinfo{year}{2020}),
  \bibinfo{pages}{1--18}.
\newblock


\bibitem[\protect\citeauthoryear{Nyhan and Reifler}{Nyhan and Reifler}{2010}]%
        {nyhan2010corrections}
\bibfield{author}{\bibinfo{person}{Brendan Nyhan} {and} \bibinfo{person}{Jason
  Reifler}.} \bibinfo{year}{2010}\natexlab{}.
\newblock \showarticletitle{When corrections fail: The persistence of political
  misperceptions}.
\newblock \bibinfo{journal}{\emph{Political Behavior}} \bibinfo{volume}{32},
  \bibinfo{number}{2} (\bibinfo{year}{2010}), \bibinfo{pages}{303--330}.
\newblock


\bibitem[\protect\citeauthoryear{Paris and Donovan}{Paris and Donovan}{2019}]%
        {paris2019deepfakes}
\bibfield{author}{\bibinfo{person}{Britt Paris} {and} \bibinfo{person}{Joan
  Donovan}.} \bibinfo{year}{2019}\natexlab{}.
\newblock \showarticletitle{Deepfakes and Cheap Fakes}.
\newblock \bibinfo{journal}{\emph{United States of America: Data \& Society}}
  (\bibinfo{year}{2019}).
\newblock


\bibitem[\protect\citeauthoryear{Pennycook, Bear, Collins, and Rand}{Pennycook
  et~al\mbox{.}}{2020}]%
        {pennycook2020implied}
\bibfield{author}{\bibinfo{person}{Gordon Pennycook}, \bibinfo{person}{Adam
  Bear}, \bibinfo{person}{Evan~T Collins}, {and} \bibinfo{person}{David~G
  Rand}.} \bibinfo{year}{2020}\natexlab{}.
\newblock \showarticletitle{The implied truth effect: Attaching warnings to a
  subset of fake news headlines increases perceived accuracy of headlines
  without warnings}.
\newblock \bibinfo{journal}{\emph{Management Science}} (\bibinfo{year}{2020}).
\newblock


\bibitem[\protect\citeauthoryear{Rosen}{Rosen}{2020}]%
        {rosen_2020}
\bibfield{author}{\bibinfo{person}{Guy Rosen}.}
  \bibinfo{year}{2020}\natexlab{}.
\newblock \bibinfo{title}{An Update on Our Work to Keep People Informed and
  Limit Misinformation About COVID-19}.
\newblock
\newblock
\urldef\tempurl%
\url{https://about.fb.com/news/2020/04/covid-19-misinfo-update/}
\showURL{%
\tempurl}


\bibitem[\protect\citeauthoryear{Saltz}{Saltz}{2020}]%
        {DVN/QAMZPO_2020}
\bibfield{author}{\bibinfo{person}{Emily Saltz}.}
  \bibinfo{year}{2020}\natexlab{}.
\newblock \bibinfo{title}{{Replication Data for: Encounters with Visual
  Misinformation and Labels Across Platforms}}.
\newblock
\newblock
\urldef\tempurl%
\url{https://doi.org/10.7910/DVN/QAMZPO}
\showDOI{\tempurl}


\bibitem[\protect\citeauthoryear{Saltz, Noel, Leibowicz, Wardle, and
  Gregory}{Saltz et~al\mbox{.}}{2020}]%
        {saltz_noel_leibowicz_wardle_gregory_2020}
\bibfield{author}{\bibinfo{person}{Emily Saltz}, \bibinfo{person}{Pedro Noel},
  \bibinfo{person}{Claire Leibowicz}, \bibinfo{person}{Claire Wardle}, {and}
  \bibinfo{person}{Sam Gregory}.} \bibinfo{year}{2020}\natexlab{}.
\newblock \bibinfo{title}{5 Urgent Considerations for the Automated
  Categorization of Manipulated Media}.
\newblock
\newblock
\urldef\tempurl%
\url{https://medium.com/partnership-on-ai/5-urgent-considerations-for-the-automated-categorization-of-manipulated-media-8fad982b2db0}
\showURL{%
\tempurl}


\bibitem[\protect\citeauthoryear{Samek}{Samek}{2018}]%
        {samek_2018}
\bibfield{author}{\bibinfo{person}{Geoff Samek}.}
  \bibinfo{year}{2018}\natexlab{}.
\newblock \bibinfo{title}{Greater transparency for users around news
  broadcasters}.
\newblock
\newblock
\urldef\tempurl%
\url{https://blog.youtube/news-and-events/greater-transparency-for-users-around}
\showURL{%
\tempurl}


\bibitem[\protect\citeauthoryear{Shearer and Grieco}{Shearer and
  Grieco}{2019}]%
        {shearer2019americans}
\bibfield{author}{\bibinfo{person}{Elisa Shearer} {and}
  \bibinfo{person}{Elizabeth Grieco}.} \bibinfo{year}{2019}\natexlab{}.
\newblock \showarticletitle{Americans are wary of the role social media sites
  play in delivering the news}.
\newblock \bibinfo{journal}{\emph{Pew Research Center}}  \bibinfo{volume}{2}
  (\bibinfo{year}{2019}).
\newblock


\bibitem[\protect\citeauthoryear{Shen, Kasra, Pan, Bassett, Malloch, and
  O’Brien}{Shen et~al\mbox{.}}{2019}]%
        {doi:10.1177/1461444818799526}
\bibfield{author}{\bibinfo{person}{Cuihua Shen}, \bibinfo{person}{Mona Kasra},
  \bibinfo{person}{Wenjing Pan}, \bibinfo{person}{Grace~A Bassett},
  \bibinfo{person}{Yining Malloch}, {and} \bibinfo{person}{James~F O’Brien}.}
  \bibinfo{year}{2019}\natexlab{}.
\newblock \showarticletitle{Fake images: The effects of source, intermediary,
  and digital media literacy on contextual assessment of image credibility
  online}.
\newblock \bibinfo{journal}{\emph{New Media \& Society}} \bibinfo{volume}{21},
  \bibinfo{number}{2} (\bibinfo{year}{2019}), \bibinfo{pages}{438--463}.
\newblock
\urldef\tempurl%
\url{https://doi.org/10.1177/1461444818799526}
\showDOI{\tempurl}
\showeprint{https://doi.org/10.1177/1461444818799526}


\bibitem[\protect\citeauthoryear{Support}{Support}{2020}]%
        {support_2020}
\bibfield{author}{\bibinfo{person}{Twitter Support}.}
  \bibinfo{year}{2020}\natexlab{}.
\newblock \bibinfo{title}{New labels for government and state-affiliated media
  accounts}.
\newblock
\newblock
\urldef\tempurl%
\url{https://blog.twitter.com/en_us/topics/product/2020/new-labels-for-government-and-state-affiliated-media-accounts.html}
\showURL{%
\tempurl}


\bibitem[\protect\citeauthoryear{Swire-Thompson, DeGutis, and
  Lazer}{Swire-Thompson et~al\mbox{.}}{2020}]%
        {swire2020searching}
\bibfield{author}{\bibinfo{person}{Briony Swire-Thompson},
  \bibinfo{person}{Joseph DeGutis}, {and} \bibinfo{person}{David Lazer}.}
  \bibinfo{year}{2020}\natexlab{}.
\newblock \showarticletitle{Searching for the backfire effect: Measurement and
  design considerations}.
\newblock  (\bibinfo{year}{2020}).
\newblock


\bibitem[\protect\citeauthoryear{Torabi~Asr and Taboada}{Torabi~Asr and
  Taboada}{2019}]%
        {torabi2019big}
\bibfield{author}{\bibinfo{person}{Fatemeh Torabi~Asr} {and}
  \bibinfo{person}{Maite Taboada}.} \bibinfo{year}{2019}\natexlab{}.
\newblock \showarticletitle{Big Data and quality data for fake news and
  misinformation detection}.
\newblock \bibinfo{journal}{\emph{Big Data \& Society}} \bibinfo{volume}{6},
  \bibinfo{number}{1} (\bibinfo{year}{2019}),
  \bibinfo{pages}{2053951719843310}.
\newblock


\bibitem[\protect\citeauthoryear{Winkielman and Gogolushko}{Winkielman and
  Gogolushko}{2018}]%
        {winkielman2018influence}
\bibfield{author}{\bibinfo{person}{Piotr Winkielman} {and}
  \bibinfo{person}{Yekaterina Gogolushko}.} \bibinfo{year}{2018}\natexlab{}.
\newblock \showarticletitle{Influence of suboptimally and optimally presented
  affective pictures and words on consumption-related behavior}.
\newblock \bibinfo{journal}{\emph{Frontiers in psychology}}
  \bibinfo{volume}{8} (\bibinfo{year}{2018}), \bibinfo{pages}{2261}.
\newblock


\end{thebibliography}

\appendix

\end{document}